\documentclass{emulateapj}
\usepackage{epsfig}

\def\spose#1{\hbox to 0pt{#1\hss}}
\def\lax{$\mathrel{\spose{\lower 3pt\hbox{$\mathchar"218$}}
     \raise 2.0pt\hbox{$\mathchar"13C$}}$}
\def\gax{$\mathrel{\spose{\lower 3pt\hbox{$\mathchar"218$}}
     \raise 2.0pt\hbox{$\mathchar"13E$}}$}

\newcommand{\cyg}{V404 Cyg}
\newcommand{\xmm}{{\it XMM-Newton}}
\newcommand{\asca}{{\it ASCA}}
\newcommand{\rosat}{{\it ROSAT}}
\newcommand{\sax}{{\it BeppoSAX}}
\newcommand{\chandra}{{\it Chandra}}
\newcommand{\lum}{\thinspace\hbox{$\hbox{erg}\thinspace\hbox{s}^{-1}$}}

\begin{document}

\title{The Spectrum of the Black Hole X-ray Nova V404 Cygni in Quiescence as Measured by {\it XMM-Newton}}

\author{Charles K. ~Bradley\altaffilmark{1}, Robert I. ~Hynes\altaffilmark{1}, Albert K. H. ~Kong\altaffilmark{2}, C. A. ~Haswell\altaffilmark{3}, J.
~Casares\altaffilmark{4}, E. ~Gallo\altaffilmark{5}}
\altaffiltext{1}{Department of Physics and Astronomy, Louisiana State University, Baton Rouge, LA, 70803}
\altaffiltext{2}{Kavli Institute for Astrophysics and Space Research, Massachusetts Institute of Technology, 77 Massachusetts Avenue, Cambridge, MA, 02139}
\altaffiltext{3}{The Open University, Walton Hall, Milton Keynes, MK7 6AA, UK}
\altaffiltext{4}{Insituto de Astrofisica de Canarias, Via Lactea, 38200 La Laguna, Tenerife, Spain}
\altaffiltext{5}{University of California, Santa Barbara, CA, 93106}

\begin{abstract}

We present \xmm\ observations of the black hole X-ray nova \cyg\ in quiescence. Its quiescent spectrum can be best fitted by a simple power-law 
with slope $\Gamma \sim 2$. The spectra are consistent with that expected for the advection-dominated accretion flow (ADAF). \cyg\ was roughly equal 
in luminosity compared to the previous observation of {\it Chandra}. We see variability of a factor of 4 during the observation. We find no evidence for
the presence of fluorescent or H-like/He-like iron emission, with upper limits of 52 eV and 110 eV respectively. The limit on the fluorescent emission is improved by a
factor of 15 over the previous estimate, and the restriction on H-like/He-like emission is lower than predicted from models by a factor of roughly 2.

\end{abstract}


\section{Introduction}

Low Mass X-ray Binaries (LMXB) are binary systems with a black hole or neutron star primary and usually a late-type main sequence or evolved donor star (Remillard \& McClintock
2006). The donor is filling its Roche Lobe and therefore transfers mass to the primary component via an accretion disk. This material within the disk emits light at all wavelengths, but is generally very faint 
 while in quiescence, however it can become very bright in X-rays and highly variable during rare outbursts, making these objects very interesting for study in the area of accretion physics. X-ray observations of these
objects only tell a portion of the story, as their orbital periods and masses are determined by optical observation of their counterparts. These orbital periods typically range
from tens of minutes to days.

Several sources of quiescent X-ray emission have been discussed for black hole X-ray binaries (BHXBs). It has been suggested that advection-dominated
accretion flows (ADAF) can explain the optical and X-ray emission and the X-ray spectra seen in quiescence (see, e.g., Narayan et al. 2002). \cyg\ has been used
as the primary testbed for many of these models because of its accessibility with X-ray observatories.  These flows also
provide a natural explanation for the luminosity difference between black hole and neutron star systems (Garcia et al. 2001). The radiatively inefficient
accreting material must impact the surface of a neutron star and be reprocessed, whereas in the black hole case, it simply
disappears beyond the event horizon. It has also been proposed by Fender, Gallo, \& Jonker (2003) that the system could go into 
a ``jet-dominated" state at such low accretion rates where a majority of the energy is released in the outflow instead of the ADAF. It has also been suggested that emission from a disk plus an
optically thin corona could explain the quiescent emission and that the standard disk + corona model should not be
discarded outright (Nayakshin \& Svensson 2001). A hybrid jet/ADAF model has been suggested by Yuan et al. (2005) applied to XTE J1118+480 that accounts for the spectral
and timing features of that system. A similar model by Malzac et al. (2004) to the same system has also been proposed that reproduces the timing features.

\cyg\ was originally discovered by {\it Ginga} during outburst in May 1989 (Makino et al 1989). Optical observation of \cyg\ determined a mass function of
$f(M)=6.08\pm0.06$\,M$_{\odot}$(Casares \& Charles 1994, Casares, Charles, \& Naylor 1992) and an orbital period of 6.47d. Combined with limits on the inclination ($i=56^{\circ}$)
we deduce a black hole mass of approximately 12 M$_{\odot}$ (Shahbaz et al. 1994), orbited by a K0IV star of $\sim$ 0.7 M$_{\odot}$. Of the stellar-mass black
hole population, \cyg\ is the X-ray brightest and most accessible object in quiescence. \cyg\ is known to vary in X-rays (Wagner et al. 1994, Kong et al. 2002) by a factor of 2
in a timescale of just 30 minutes, and by a factor of 10 within 12 hours in quiescence.

The origin of the quiescent variability remains uncertain, however, with possibilities ranging from magnetic reconnection events, to local
changes in the accretion rate. Optical and X-ray variations in \cyg\ in quiescence have been
shown to be fairly well correlated by Hynes et al. (2004). They showed that
variability is mirrored well by H$\alpha$ and to a lesser extent by the optical continuum.This correlated variability, generally attributed to irradiation of the outer cool(er)
disk, is commonly seen in bright X-ray states for both neutron star and black hole systems (Hynes 2005).

Here we report the analysis of {\it XMM-Newton} spectra of the brightest quiescent BHXB observed, 
V404 Cyg. This is the highest quality X-ray spectrum of a quiescent BHXB yet obtained. We briefly describe previous quiescent observations of this source in \S\,2. 
In \S\,3 we outline the analysis procedure and report results in \S\,4 and \S\,5. They are then discussed in \S\,6.

\begin{table*}[t]
\begin{center}
\caption {Previous Quiescent Observations}
\begin{tabular}{llcccccc}
\hline \hline
Source&Date& Instrument & $N_H$& $\Gamma$ & Luminosity & Distance &References\\
    &&            & ($10^{22}$ cm$^{-2}$) & &($10^{33}$ \lum)  & (kpc)&\\
\hline
\cyg\ &1992 Nov & \rosat & 2.29 \tablenotemark{a} & 6 \tablenotemark{a} & 8.1 (0.1--2.4keV)& 3.5 &1\\
&&  & 2.1 \tablenotemark{a} & $4.0^{+1.9}_{-1.5}$ & 1.1 (0.7--2.4 keV)& &2 \\
&1994 May & \asca     & $1.1^{+0.3}_{-0.4}$      & $2.1^{+0.5}_{-0.3}$    & 1.20  (1--10 keV) &     &3\\
&1996 Sept & \sax     & 1.0 (fixed)              & $1.9^{+0.6}_{-0.3}$    & 1.04 (1--10keV) &     &4\\
&2000 Apr  & \chandra & $0.69\pm0.08$   & $1.81\pm0.14$            &  2.07 (0.3--7keV) & 3.5 &  5 \\
&2003 Jul  & \chandra & $0.75\pm0.07$   & $2.17^{+0.11}_{-0.15}$ &  1.10 (0.3--7keV) & 3.5 &  6 \\
&2006 Nov  & \xmm     & $0.88\pm0.6$    & 2.09 $\pm$ 0.08        &  1.07 (0.3--10keV)& 3.5 &  7 \\
\hline
\end{tabular}
\end{center}
\tablecomments{(1) Wagner et al. 1994; (2) Narayan et al. 1996; (3) Narayan et
al. 1997a; (4) Campana et al. 2001; (5) Kong et al. 2002; (6) Hynes et al. (in prep), (7) this result}
\tablenotetext{a}{denotes that uncertainty was not given}
\end{table*}

\section{Previous observations of V404 Cyg}

This bright quiescent BHXB has been observed by {\it ROSAT}, {\it
ASCA}, {\it BEPPO-SAX}, and twice by {\it Chandra}. In general, its
spectrum has been well fitted by a simple power-law model with photon
index $\Gamma\sim 2$ with varying estimates of $N_{\rm H}$ in the range $\sim
0.7-2.3 \times 10^{22}$ cm$^{-2}$ and a luminosity of $\sim 1.0 \times
10^{33}$ erg/s (see Table 1; Narayan et al. 1997, Kong et
al. 2002).  Early estimates of the spectrum were consistent with those
derived by {\it Chandra}, when one considers that the {\it ROSAT} data
are based on energies below 2.5\,keV in which range errors on $N_{\rm H}$ and $\Gamma$ are strongly positively correlated.
Previous upper limits on the equivalent width of the 6.4
keV iron line emission were found to be $\sim 800$ eV (Kong et
al. 2002) and were too high to constrain spectral models that included
line emission.

\section{XMM observations and data reduction}

\cyg\ was observed by \xmm\ on 2005 November 8-9. The observation lasted 40ks. The three EPIC cameras were all active during the observation: pn, EMOS1, and EMOS2. The three
cameras were operated with a medium filter in Imaging Full Window Mode. Total background corrected counts from the three cameras are given in Table 2. 
For this paper we analyzed standard pipeline-processed event files with SAS v6.2.0. In order to study the temporal behavior of \cyg\ we accumulated its pn lightcurve in the energy range 0.3-10 keV, 
using an extraction radius of 6 pixels (Figures 1-2). EMOS camera analysis is restricted to the 0.3-7 keV range, also with an extraction radius of 6 pixels. Corresponding backgrounds were taken from the same chip as the source but far enough away from it so as to not be affected by the wings
of the point spread function, with an extraction radius of 16 pixels in pn and in MOS. These lightcurves show variability of a factor of 4 in count rate within 10ks, and a large
amplitude flare is detected at the very beginning of the run with a duration of about thirty minutes.

\begin{figure}
\epsscale{0.7}
\begin{center}
\includegraphics[scale=0.5]{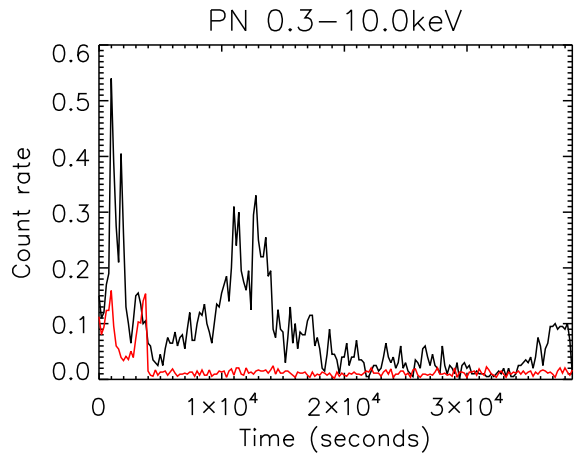}
\caption{PN camera lightcurve, source and background, binned at 200s}
\label{fig1}
\end{center}
\end{figure}

\begin{figure}
\epsscale{0.7}
\includegraphics[scale=0.5]{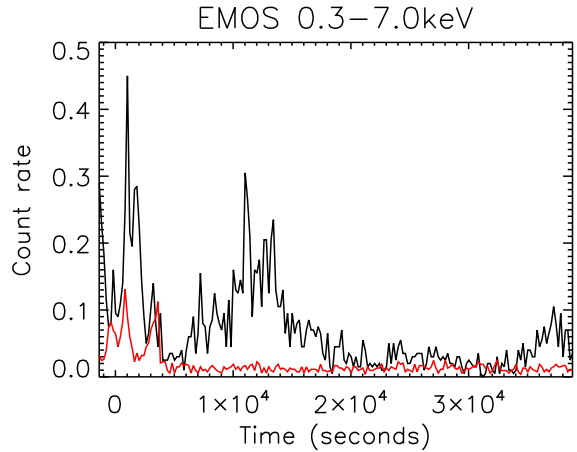}
\caption{Summed EMOS lightcurves with background, binned at 200s} 
\label{fig2}
\end{figure}

The pn pixels are larger than the EMOS, so no multi-pixel events are expected by the soft-proton background, therefore a standard Pattern 0 cleaning was used for the lightcurve.
Standard pn filtering was accomplished with XMMEA-EP and EMOS filtering with XMMEA-EM. We measured a source count rate of 0.035 $s^{-1}$ for pn and an average of 0.012 $s^{-1}$
each for the two EMOS.

\section{Continuum spectral analysis}

Spectra were extracted with XMMselect and were analyzed with XSPEC v12.2 \footnote{http://heasarc.gsfc.nasa.gov/docs/xanadu/xspec/index.html} and we report those results here. 
Spectral extraction was done with XMMEA-EP and Pattern 4. EMOS extraction was done likewise with XMMEA-EM and with Pattern 12. Bin sizes in extraction were 5 counts/channel
for pn and 15 for the EMOS cameras, as suggested in the \xmm\ Users Guide \footnote{http://xmm.esac.esa.int/external/xmm\_user\_support/documentation/sas\_usg/USG/USG.html}for optimal use of the standard response matrices. Single and double events were included under a Pattern
4 extraction for pn, and all valid patterns were kept for EMOS. Response files were made using standard SAS tools to check against the standard version, with consistent results
for the spectral fits from both. Further spectral analysis was done between 0.3-10 keV for pn and 0.3-7 keV with MOS.

We have chosen to use $\chi^2$ Churazov-weighted 
statistics (Churazov 1996) in our analysis, since we feel these to be more applicable when line 
diagnostics are to be done. CASH statistics (Cash 1979) were used as a check. For this we used lightly binned data (5cts/bin) for $\chi^2$ statistical fits. We then fit the data with 
many single-component spectral models such as power-law, thermal bremsstrahlung, Raymond-Smith (1977), and blackbody, including interstellar absorption. 
A broken power-law model was also tried with data from the onboard pn camera, but fits were poor to EMOS1 and EMOS2 data. 
Fits to many other single and multiple component models within XSPEC were tried (comptonized blackbody, blackbody with power-law, etc.), 
but in those cases the error bars on fitted parameters were unreasonably large or the fit was too poor. The best-fit 
parameters for each of the three cameras used, by both CASH and $\chi^2$ statistics, are consistent between the two fitting methods and only $\chi^2$ results are reported
herein. The quoted errors on the derived model parameters correspond to a 90\% confidence level.

Only two of the models reported give statistically acceptable fits to the data ($\chi^2/\nu$ \lax 1). 
The power-law model provides the best fit to the data, and yields parameters consistent with previous observations 
(e.g. $\Gamma \sim 2.0$; see Table 2), and was then used for line diagnostics. Bremsstrahlung is also acceptable, while Raymond-Smith and blackbody model fits are rejected.
This best fitting model is shown in Figures 3 and 4.

\begin{table*}[t]
\begin{center}
\caption{Best--fit parameters for power--law and bremsstrahlung models}\label{fit}
\footnotesize{
\begin{tabular}{ccccccccc}
\hline
Instrument & Model          & $N_{H}$                    & Photon & kT  & Normalization at 1 keV            & $\chi^{2}_{\nu}$/dof   & $f_{X}$ (0.3--10keV)\tablenotemark{a} & counts     \\

           &                & (10$^{22}$ cm$^{-2}$) & Index  &     &(ph keV$^{-1}$ cm$^{-2}$ s$^{-1}$) &                        & (erg cm$^{-2}$ s$^{-1}$) &        \\\hline

pn         & PL             & 0.86 $\pm$ 0.7         & 2.10 $\pm$ 0.10 & --                     & (1.98 $\pm$ 0.2) $\times 10^{-4}$     & 0.75/115 & 1.04 $\times 10^{-12}$  & 1404 \\

           & bremsstrahlung & $0.69^{+0.05}_{-0.02}$ & --              & $4.88^{+0.73}_{-0.6}$  & (1.67 $\pm$ 0.2) $\times 10^{-4}$     & 0.77/115 & 7.47 $\times 10^{-13}$& \\

EMOS1      & PL             & 0.859 $\pm$ 1.2        & 2.04 $\pm$ 0.18 & --                     & (1.99 $\pm$ 0.4) $\times 10^{-4}$     & 0.96/48  & 1.06 $\times 10^{-12}$  & 497      \\

           & bremsstrahlung & $0.67^{+0.1}_{-0.08}$  & --              & $5.40^{+1.9}_{-1.25}$  & (1.72 $\pm$ 0.2) $\times 10^{-4}$     & 1.01/48  & 7.31 $\times 10^{-13}$&              \\

EMOS2      & PL             & 0.927 $\pm$ 1.3        & 2.06 $\pm$ 0.18 & --                     & (2.07 $\pm$ 0.4) $\times 10^{-4}$     & 0.92/47  & 1.12 $\times 10^{-12}$  & 519      \\ 

           & bremsstrahlung & $0.75^{+0.1}_{-0.09}$  & --              & $4.91^{+1.72}_{-1.09}$ & (1.80 $\pm$ 0.3) $\times 10^{-4}$     & 1.05/47  & 7.37 $\times 10^{-13}$&              \\

JOINT FIT  & PL             & 0.88 $\pm$ 0.6         & 2.09 $\pm$ 0.08 & --                     & (2.02 $\pm$ 0.2) $\times 10^{-4}$     & 0.86/216 & 1.08 $\times 10^{-12}$&\\
           & bremsstrahlung & 0.70 $\pm$ 0.4         & --              & $4.93^{+0.59}_{-0.49}$  & (1.72 $\pm$ 0.1) $\times 10^{-4}$     & 0.89/216 & 7.81 $\times 10^{-13}$ & \\ \hline 
\end{tabular}}
\end{center}
\tablecomments{Errors are at 90\% c.l. for a single interesting parameter;}
\tablenotetext{a}{unabsorbed flux; flux is in 0.3-7 keV range for MOS}
\end{table*}

The pn and EMOS spectra give consistent fit parameters with quoted uncertainties (see Table 2). Combined data fits restrict $\Gamma \sim 2.09$, $N_{\rm H} = 0.7-0.9 \cdot 10^{22}$ cm$^{-2}$ for the two models considered, but with a
luminosity for the two of $L_{x} \sim 1.0 \times 10^{33}$ erg s$^{-1}$, consistent with previous results.
In the following analysis, we aim to constrain the possible presence of the iron line and consider only the pn spectra, due to the higher statistics at high energies.

\begin{figure}
\centering
\resizebox{\hsize}{!}{\includegraphics[angle=-90, clip=true]{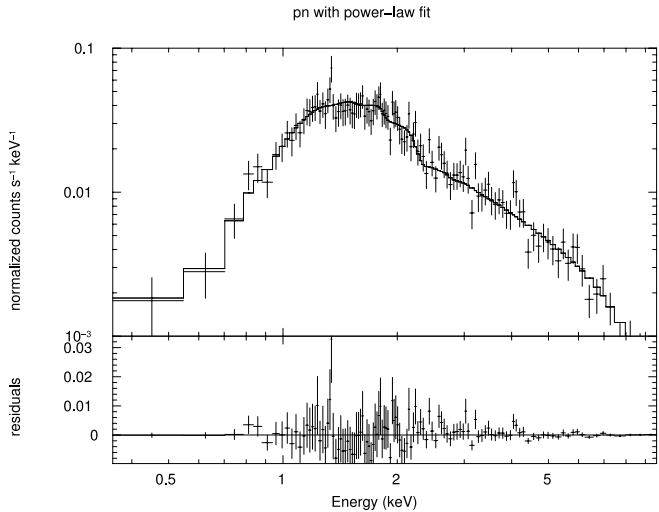}}
\caption{Comparison of the PN spectrum with the best--fit power--law model. In the lower panel the residuals between data and model are shown, binned for visual clarity}
\label{fig4}
\end{figure}

\begin{figure}  
\begin{center}
 {\label{fig_01}\includegraphics[scale=0.3, angle=-90]{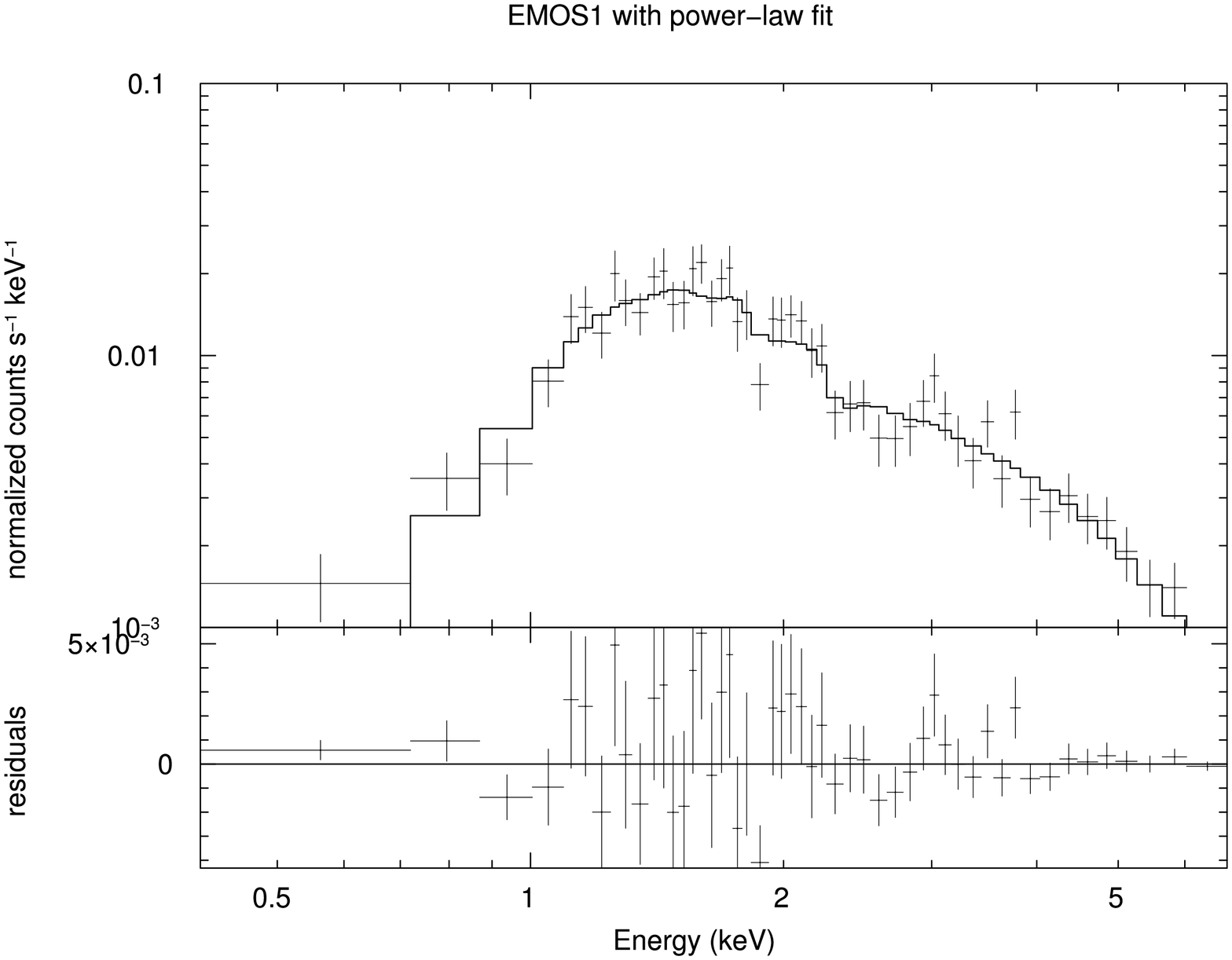}}
 {\label{fig_02}\includegraphics[scale=0.3, angle=-90]{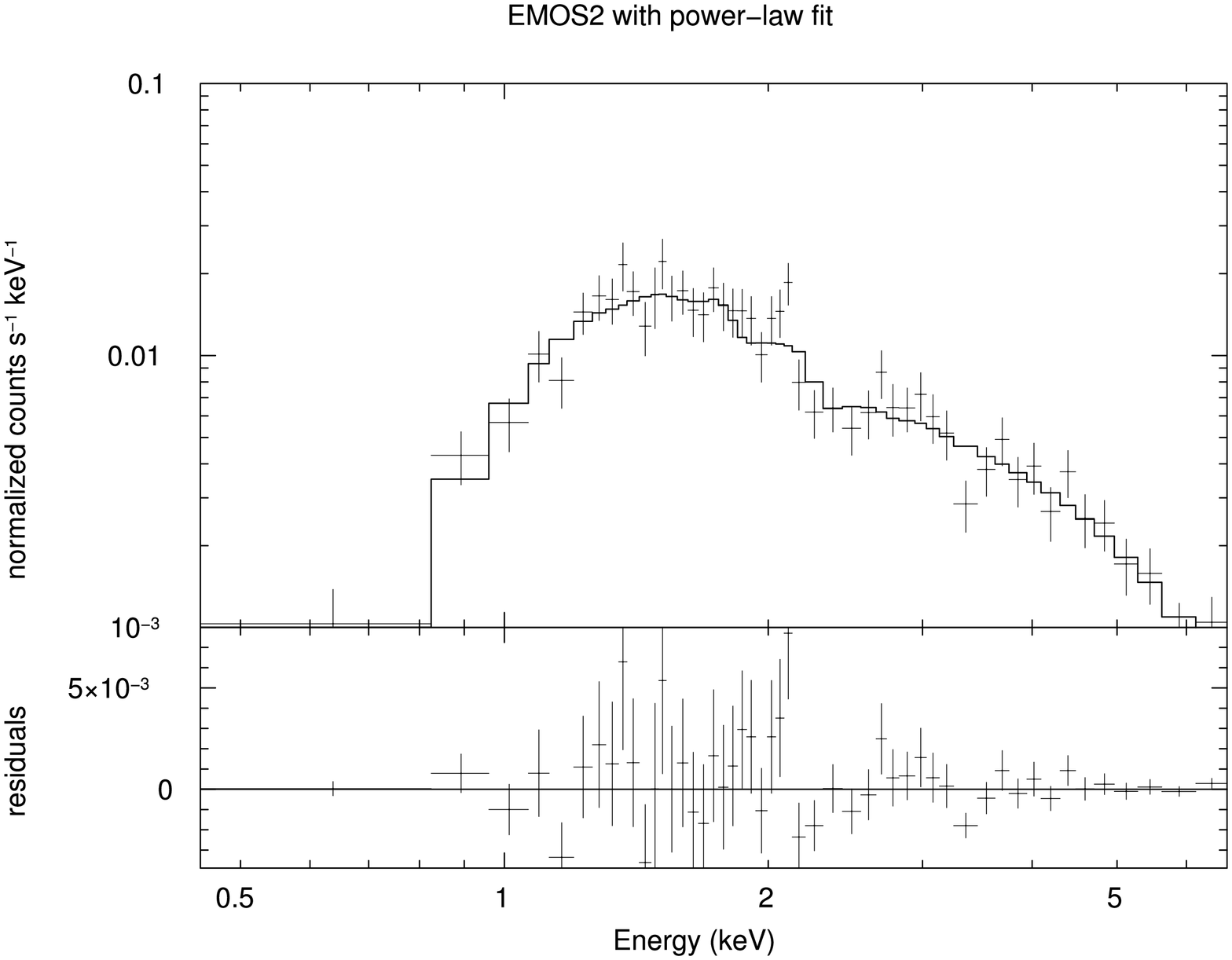}}
\caption{EMOS 1 and 2 spectrum, 0.3--7 keV, fitted with the power-law model and showing residuals, binned for visual clarity}
\end{center}
\label{fig5}
\end{figure}

\section{Line diagnostics}

\subsection{Iron K$\alpha$ lines in XRBs}

Iron K$\alpha$ emission features have been detected in other X-ray binaries before, usually but not exclusively (Oosterbroek et al. 2001) in BH transients. A few 
recent examples of stellar-mass BH cases are XTE J1650-500 (Miller et al. 2002), SAX J1711.6-3808 (In't Zand et al. 2002b),
 XTE J2012+381 (Campana et al. 2002) and GRS 1915+105 (Martocchia et al. 2002). However these detections are generally weak or broad lines.
 See also Miller (2006) for a recent review.

Before searching for an iron K$\alpha$ line in our data, it must be stressed that detection of broad lines represents an
observational challenge, due to limited statistics and uncertainty in
the shape of the underlying continuum.  The lines are easier to detect
if the disk is truncated at larger radii and/or the radial emissivity
profile is not centrally concentrated, so that the line is not broad
and can be distinguished from the continuum. On the other hand, if the
black hole is rapidly spinning so that a standard cold disk can extend
to small radii and the profile is centrally concentrated (as it should
be), the line becomes so broad that its low contrast against the
continuum makes it a challenge for detection even for \xmm\ (Fabian \&
Miniutti 2005). In the latter case, the line would only be detected
easily if the system has a super-solar abundance and/or a large
reflective component.  Fortunately, in quiescence the disk is not
expected to extend close to the black hole, and thus we do not expect a
broad line, at least in the dominant paradigm of an evaporated inner
disk.  While, Nayakshin \& Svennson (2001) dispute that the disk
truncates in this way, they also do not expect central concentration
of the X-ray emitting region. In their model, most of the X-ray
emission comes from a corona above a thin disk at large radii; hence a
narrow line would again be expected.  Given the limited statistical
quality of our data, we therefore confine our search to narrow lines.
We should keep in mind, however, that if the line were much broader
than these expectations then we are unlikely to detect it without a
much higher quality spectrum and coverage of the high energy continuum.

\subsection{Observational contraints on V404 Cygni}
Our spectral analysis of \cyg\ shows no evidence for a narrow 6.4 keV iron line. From adding a gaussian component to the best-fitting model,
(fixed width 0.1 keV, corresponding to instrumental resolution) we establish an upper limit of $\sim 52$ eV for this line (90 \% confidence), well below all previous estimates ($\sim 800$ eV, Kong et al. 2002). We then
repeated this analysis using CASH statistics, and find the
results to be consistent.

Prompted by predictions of emission in ADAF-like flows by Narayan \& Raymond (1999), we then test-fitted a 6.7 keV gaussian (from He-like Fe XXV) to the spectrum, and detect no 
significant emission there either with an upper limit of $\sim 110$ eV(0.1 keV fixed width; 90 \% confidence) on the feature. The total group of Fe line widths in their paper sums
to roughly 230 eV, a factor of about 2 above our results. A close-up of the 5-10 keV region of the spectrum is presented in Fig. 5. 

There are some features seen in the spectrum of \cyg\ and possibly within previous results (Hynes et. al in prep) that may indicate the presence
of line emission at just above 4 keV. This line corresponds to a line predicted from Ca XIX/XX in Narayan \& Raymond (1999). We again added in a gaussian line
component (allowing the energy to vary) to the power-law fit at an energy of 4.08 keV (Ca XX Ly$\alpha$), and find a fitted equivalent width of $\sim 117$ eV, well
above the 2.7 eV equivalent width predicted in their paper, for a model with wind outflows. The 90\% confidence range on the equivalent width is from 4.0-210 eV.

\begin{figure}[h]
\begin{center}
\includegraphics[scale=0.3, angle=0]{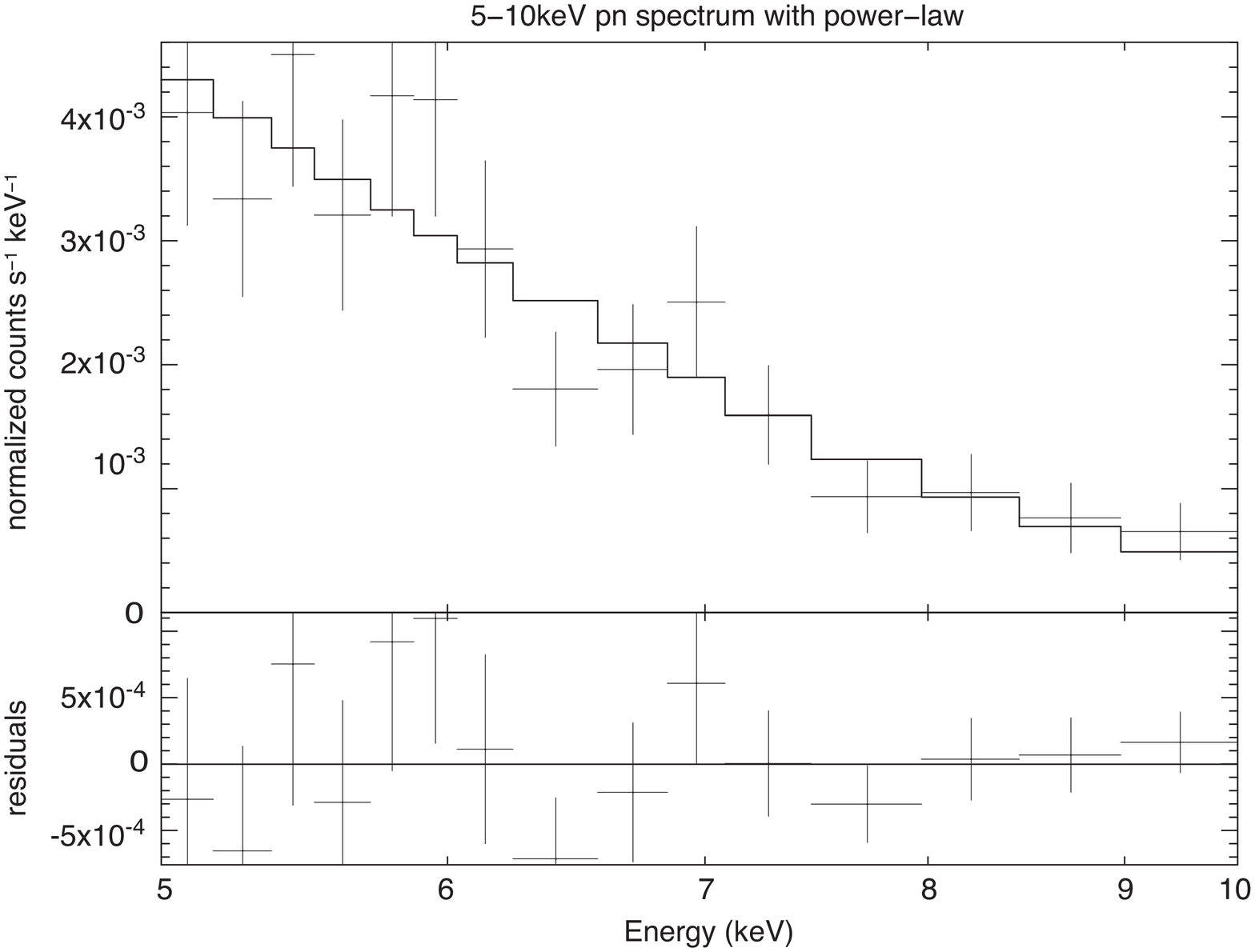}
\caption{pn spectrum in energy range of 5-10 keV, with powerlaw model overlaid}
\end{center}
\label{fig_03}
\end{figure}

\begin{figure}[h]
\begin{center}
\includegraphics[scale=0.3, angle=-90]{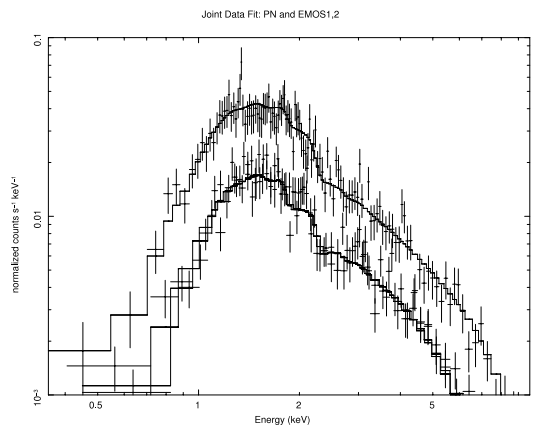}
\caption{pn and EMOS spectra joint fit to power-law model}
\end{center}
\label{fig_04}
\end{figure}

\section{Discussion}

\subsection{Iron fluorescent emission}

One would expect 6.4 keV fluorescent emission from the presence of
cold material near the X-ray emitting region, or an extended emission
region above the disk that illuminates cold material below.
For typical inclination angles, (e.g., $i=56^{\circ}$ for \cyg\ ) the absence of such a feature constrains the reflection fraction and
suggests that there is no cold matter near the X-ray emission region, or at least that such material does not subtend a large solid angle as seen by the X-ray source.
George and Fabian (1991) considered a model in which the X-ray source
was located at a height above the central region of a standard disk,
for a semi-isotropically and isotropically illuminated cold slab
(without rotation), and calculated equivalent widths of the K$\alpha$
line for varying inclinations (relative to the observer) and incident
power-law indices. Their predicted equivalent width for the power-law index and inclination ($i=56^{\circ}$) of \cyg\ for the 6.4 keV
feature was
$\sim 110-120$ eV, a factor of 2 higher than our upper limit (see
Figure 14 of their paper).  The lack of the a narrow 6.4\,keV line is
then inconsistent with the simplest geometry of an X-ray source
sandwiching a cold, slowly rotating slab.  In particular our
observations would seem to disagree with expectations of the Nayakshin
\& Svensson model (2001) for this system because their emission comes
from large radii, where there should be a cold thin disk, so
we would expect a narrow line to be present. The Nayakshin \& Svensson
model has other difficulties as well, such as higher luminosity
predictions than what are observed for quiescent BH systems. At these
luminosities we would expect some signature of 6.4\,keV emission. 
 
The discussion above has, of course, neglected rotation, The 0.1\,keV
fixed width adopted above corresponds to an unresolved line from a
region extending no closer than $\sim1000_{\rm sch}$ to the compact
object.  This is smaller than typically assumed in modeling advective
flows, but we cannot observationally exclude the possibility of broad
line emission from a region closer to the black hole.

\subsection{6.7 keV emission}

We would expect Fe XXV K$\alpha$ emission at an energy of 6.7 keV from a hot coronal/ADAF flow with high collisional rates.
The lack of any such 6.7 keV line emission may point back to earlier ADAF models of Comptonized thermal synchotron emission within a small region close to the black hole. Such 
a small emitting region would also fail to properly illuminate the disk surface to achieve significant 6.4 keV emission. 

In particular, one of the ADAF models with moderately strong outflows presented in Narayan \& Raymond (1999) is ruled out. This paper specifically addresses X-ray spectra of
systems thought to contain an ADAF flow. Their spectral model consisted of what is basically a Raymond-Smith (1977) plasma with an additional Compton component added in. They follow a method outlined in Quataert \& Narayan (1999), and assign reasonable values to the microscopic parameters of the system,
specifically the viscosity parameter and plasma parameter, which seem to reproduce the correct spectral index otherwise, and vary the fraction of dissipated heat that goes directly
into the electrons between their wind and no-wind models. The spectral calculation starts at a radius of 100 Schwarzchild radii and proceeds outward, with the intent to target the
transition region from standard thin disk to ADAF where most of these lines would originate. They do not include synchotron emission or the effects of
Comptonization, and no Doppler smearing, assumed to be small at the radii of interest. Inner regions are assumed to be so hot that the gas is completely ionized, and here they computed the continuum emission but no line emission.

In this paper it was suggested by the authors that the equivalent width of lines correspond to the size of the ADAF region, and that for a given size the widths increase with outflow strength. Lower accretion rates would lead to 
increased bremsstrahlung emission, and equivalent widths increase. A bremsstrahlung model fit to the data is statistically acceptable, but we see no lines, the most prominent 
of which should have been at 6.7 keV. Based on their conclusions, our results would indicate that \cyg\ is a Compton-dominated system with a weak outflow, if one is present.
However the assumptions made and outlined above may be called into question. The plasma parameter, the viscosity, the accretion rate's dependence on radius, or the fraction of heat transferred to the electrons in the ADAF
could all very well be different from that assumed. They state that their model is insensitive to the values assumed for the viscosity and plasma parameters however. Their model
assumes ionization equilibrium, but they point out that departure from this will only make the line intensities larger which is in the opposite sense to that caused by
photoionization. The modeling of the corona above the outer thin disk is crude, however. The strongest X-ray line emission occurs for $T_e \sim 10^7-10^8$ K, suggesting that the source of the
X-ray emission is above this temperature. 

A simpler model of hot optically-thin plasma emission in ionization equilibrium (Raymond \& Smith 1977) appears to be statistically ruled out by the combination of our results with those previously published (Kong et al. 2002).

\subsection{Other explanations for the absence of line emission}
Absence of any iron line may indicate that the system is underabundant in iron, all the material present in the X-ray emitting region is completely ionized, or perhaps that resonant trapping is efficiently
eliminating any line emission (Matt et al. 1993, 1996). A hole in the disk filled with an ADAF would explain the absence of cold material, but we may still expect reflection.
The presence of a central ADAF should diminish the strength of any reflective feature (Gierlinski et al. 1997) but not necessarily push it to zero (Miller et al. 2002). 

The absence of any Fe line is not likely to be caused by underabundance because a strong Fe line feature with an equivalent
width of 130-150 eV was detected in outburst (Oosterbroek et al. 1996). There is little other evidence for peculiarity of any abundances in \cyg\, aside from a higher than expected lithium abundance in the companion (Casares et al. 1993, Casares \& Charles 1994, Martin et al. 1992) which may be associated with the ADAF
(Yi \& Narayan 1997), although other Li production scenarios have also been proposed (e.g. Martin et al. 1992). The possible calcium line at 4.08 keV is intriguing, although this
is not a secure detection. Ca-44 could be produced as a secondary decay product of Ti-44 (half life of $\sim 86$ years), which is produced in a supernova explosion (Theiling \& Leising 2006), however this system's supernova could not have been recent enough for an overabundance to remain, and no
overabundance is seen in the optical spectra. Other explanations for higher abundances of calcium have been proposed in the context of nucleosynthesis within thick accretion disks with
very low viscosity (Arai \& Hashimoto 1992, note Fig. 4; Mukhopadhyay \& Chakrabarti 2000, Jin, Arnett \& Chakrabarti 1989) but it would be premature to invoke such models without
a more secure detection.

\section{Conclusions}
We have presented new {\it XMM-Newton} spectroscopy of the brightest
quiescent stellar mass black hole, \cyg.  Fits to the continuum
statistically reject all fitting models considered with the exception
of power-law or bremsstrahlung emission.  We place stringent new upper
limits (52 eV, 90\,\%\ confidence) on the presence of a narrow 6.4 keV
line from neutral iron. Previous upper limits on the equivalent width
of the 6.4 keV iron line emission were found to be $\sim 800$ eV (Kong
et al. 2002), an improvement by a factor of 15.  We also place a limit
of 110 eV on ionized iron emission at 6.7 keV, roughly a factor of 2 below a 
specific model prediction.

Overall, our results are consistent with the common interpretation of
the quiescent state involving a truncated accretion disk, within which
a diffuse quasi-spherical coronal/ADAF flow exists and attempts to
cool itself by optically-thin emission.  Variations like the adiabatic
inflow/outflow (Blandford \& Begelman 1999) allow for a significant
fraction of the outer material to be lost to an outflow at large
radii, and have rotation.

The lack of strong ionized lines apparently argues against a strong
outflow/wind in the context of specific ADAF models considering winds
(Narayan \& Raymond 1999), where lower central accretion rates
correspond to an enhancement on the bremsstrahlung emission and line
widths.  It is hard to draw strong general conclusions from this
specific comparison, however.  Indeed we know from radio observations
that \cyg\ does power a synchotron emitting outflow (whether this
comes from the innermost region or not is not known) (Gallo, Fender,
\& Hynes 2005).

\acknowledgements
This work is based on observations obtained with XMM-Newton, an ESA science
mission with instruments and contributions directly funded by ESA Member States
and NASA. The authors would like to thank Juhan Frank and Chris Mauche for their helpful comments. JC acknowledges support from the Spanish Ministry 
of Science and Technology through the project AYA2002-03570. We acknowledge support from the XMM Guest Observer Program and grant NNG06GB64G. CB acknowledges support from the
LaSpace GSRA.

\end{document}